\documentclass[11pt]{article}
\usepackage{moriond}
\usepackage{xspace}
\usepackage{amsmath}
\def\be{\begin{equation}}
\def\ee{\end{equation}}
\def\bea{\begin{eqnarray}}
\def\eea{\end{eqnarray}}

\newcommand{\geneva}{\textsc{Geneva}\xspace}
\newcommand{\pythia}{\textsc{Pythia}\xspace}
\newcommand{\cut}{\mathrm{cut}}
\newcommand{\Aleph}{ALEPH\xspace}
\newcommand{\Opal}{OPAL\xspace}
\newcommand{\abs}[1]{\lvert#1\rvert}

\newcommand{\df}{\mathrm{d}}

\newcommand{\Tau}{\mathcal{T}}

\newcommand{\GeV}{\,\mathrm{GeV}}

\newcommand{\nn}{\nonumber}

\newcommand{\FO}{\mathrm{FO}}

\newcommand{\resum}{\mathrm{resum}}

\renewcommand{\max}{\mathrm{max}}

\newcommand{\Ecm}{E_\mathrm{cm}}

\newcommand{\Photo}{}

\begin{document}
{\flushright DESY 13-086\\}
\vspace*{4cm}
\title{COMBINING HIGHER-ORDER RESUMMATION WITH MULTIPLE\\ NLO CALCULATIONS AND PARTON SHOWERS IN THE GENEVA\\ MONTE CARLO FRAMEWORK}

\author{Simone Alioli,$^1$ \footnote{Speaker}
Christian W.~Bauer,$^1$
Calvin Berggren,$^1$
Andrew Hornig,$^2$
Frank J.~Tackmann,$^3$
Christopher K.~Vermilion,$^1$
Jonathan R.~Walsh,$^1$
Saba Zuberi$^1$}

\address{$^1$Ernest Orlando Lawrence Berkeley National Laboratory and \\ University of California, Berkeley, CA 94720, U.S.A.\\
$^2$Department of Physics, University of Washington, Seattle, WA 98195, U.S.A.\\
$^3$Theory Group, Deutsches Elektronen-Synchrotron (DESY), D-22607 Hamburg, Germany}

\maketitle\abstracts{
We discuss the \geneva Monte Carlo framework,
which combines
higher-order resummation (NNLL) of large Sudakov logarithms with multiple next-to-leading-order (NLO) matrix-element corrections and parton showering (using \pythia8) to give a complete description at the next higher perturbative accuracy in $\alpha_s$
at both small and large jet resolution scales.
Results for $e^+e^- \to $ jets compared to LEP data and $pp \to (Z/\gamma^* \to \ell^+ \ell^-)$ + jets are presented.
}

\section{Introduction}

Present and future colliders require accurate and reliable predictions of QCD effects, beyond the lowest perturbative accuracy in the strong coupling $\alpha_s$ expansion. 
For inclusive observables, such as total cross sections, the lowest perturbative accuracy is obtained via a fixed-order expansion in powers of $\alpha_s$, truncated at the leading order. To get accurate results, one is usually forced to go at least to the next higher order, i.e., NLO, or even to NNLO.   
For more exclusive observables, the presence of logarithmically enhanced contributions in certain regions of phase space requires an all-orders resummation to obtain physically meaningful results.
In this case, the proper lowest perturbative accuracy is the (N)LL resummation. In general, a description which aims to be valid across the entire phase space demands a combination of both types of corrections. 
At the lowest order, such a combination is
achieved in Monte Carlo programs by the standard merging of matrix elements with parton showers (ME/PS).~\cite{Catani:2001cc,Lonnblad:2001iq}

The \geneva framework~\cite{Alioli:2012fc} extends this to higher perturbative accuracy by including the fixed NLO corrections as well as the NNLL resummation of the jet resolution parameter, which in our case is chosen to be $N$-jettiness~\cite{Stewart:2010tn} due to its simple factorization and resummation properties.
An immediate by-product of this combination of fixed-order and resummed results for different jet multiplicities is the
merging of multiple NLO calculations, which has been the subject of several recent theoretical efforts.~\cite{Hoeche:2012yf,Frederix:2012ps,Alioli:2011nr,Platzer:2012bs,Lonnblad:2012ix} The key difference in our approach is the inclusion of higher logarithmic resummation~\footnote{The inclusion of higher logarithmic resummation has also been proposed in a subsequent work~\cite{Hamilton:2012rf} as a possible way to remove the dependency on the jet resolution scale.} 
of the jet resolution scale $\tau^\cut$, which allows us to push
it to much lower values than fixed-order perturbation theory would allow.
In this way, we can avoid the restriction $\alpha_s \ln^2\tau^\cut \ll 1$ that limits the range of applicability of
other approaches.

To provide further parton showering and hadronization, \geneva is interfaced to \pythia8.~\cite{Sjostrand:2006za} In this way, the best possible theoretical predictions in the context of fully exclusive Monte Carlo event generators can be  directly made available for experimental analyses.
\section{Theoretical framework}
We now give a brief description of our method, referring to the \geneva paper~\cite{Alioli:2012fc} for a comprehensive discussion.
We first separate the exclusive $N$-jet and inclusive $(N+1)$-jet regions,
\begin{equation}\label{eq:inclxsec}
\sigma_{\geq N}
= \int\!\df\Phi_N\, \frac{\df\sigma}{\df\Phi_N}(\Tau_N^\cut)
+ \int\!\df\Phi_{N+1}\,\frac{\df\sigma}{\df\Phi_{N+1}}(\Tau_N) \,\theta(\Tau_N > \Tau_N^\cut)
\,,\end{equation}
where $\df\sigma/\df\Phi_N(\Tau_N^\cut)$ is the fully differential $N$-jet cross section for $\Tau_N < \Tau_N^\cut$ and  $\df\sigma/\df\Phi_{N+1}(\Tau_N)$ is the fully differential
cross section for a given $N$-jettiness value $\Tau_N(\Phi_{N+1})$.
The parameter
$\Tau_N^\cut$ is a small infrared cutoff $\sim 1\,\mathrm{GeV}$, whose
dependence in the final results cancels to the required resummation order.
In the $N$-jet region, where
$\Tau_N$ is small, we then resum the logarithms of $\Tau_N/Q$, with
$Q$ some hard scale of the process.  In the $(N+1)$-jet region, at
large $\Tau_N$, we instead use a fixed-order expansion in $\alpha_s$.
To properly combine the higher fixed-order results at large $\Tau_N$
with the higher-order resummation at small $\Tau_N$, with a smooth
transition between these two regimes, we employ the
following master formulas,
\begin{align}\label{eq:introcumspec} 
\frac{\df\sigma}{\df\Phi_{N}}(\Tau_N^\cut)
&= \frac{\df\sigma^\resum}{\df\Phi_N}(\Tau_N^\cut)
+ \biggl[\frac{\df\sigma^\FO}{\df\Phi_{N}}(\Tau_N^\cut)
- \frac{\df\sigma^\resum}{\df\Phi_N}(\Tau_N^\cut)\bigg\vert_\FO \biggr]
\,, \nn \\[1ex]
\frac{\df\sigma}{\df\Phi_{N+1}}(\Tau_N)
&= \frac{\df\sigma^\FO}{\df\Phi_{N+1}}(\Tau_N)
\biggl[\frac{\df\sigma^\resum}{\df\Phi_N \df\Tau_N}\bigg/\frac{\df\sigma^\resum}{\df\Phi_N\,\df\Tau_N}\bigg\vert_\FO \biggr] 
\,,\end{align}
 where the superscript ``resum'' indicates an analytically resummed calculation and ``FO'' indicates a fixed-order calculation or expansion. 
This construction can be iterated in the case of several multiplicities.~\cite{Alioli:2012fc}
At this stage, we have explicit control of the perturbative uncertainties and are able to estimate reliably both the fixed-order and resummation uncertainties and to combine them
to provide perturbative event-by-event uncertainties.

Once the partonic \geneva events are generated according to Eq.~(\ref{eq:introcumspec}) -- or its generalization in case of more jet multiplicities --
they are fed through the \pythia8 parton shower, whose purpose it is to fill up the jets with additional emissions. To preserve the perturbative accuracy of the higher-order resummation, the shower is constrained not to change the weight of an event and to preserve its value of $\Tau_N$. (In general, this is a nontrivial constraint and can be implemented with sufficient approximation and manageable efficiency.)

Finally, we rely on the \pythia8 hadronization model to hadronize the final-state partons. No further constraints are applied in this step, since \geneva's partonic predictions do not include any nonperturbative effects.
\section{Results}
\begin{figure}[!ht]
\begin{minipage}{0.33\linewidth} \centerline{\includegraphics[width=\linewidth]{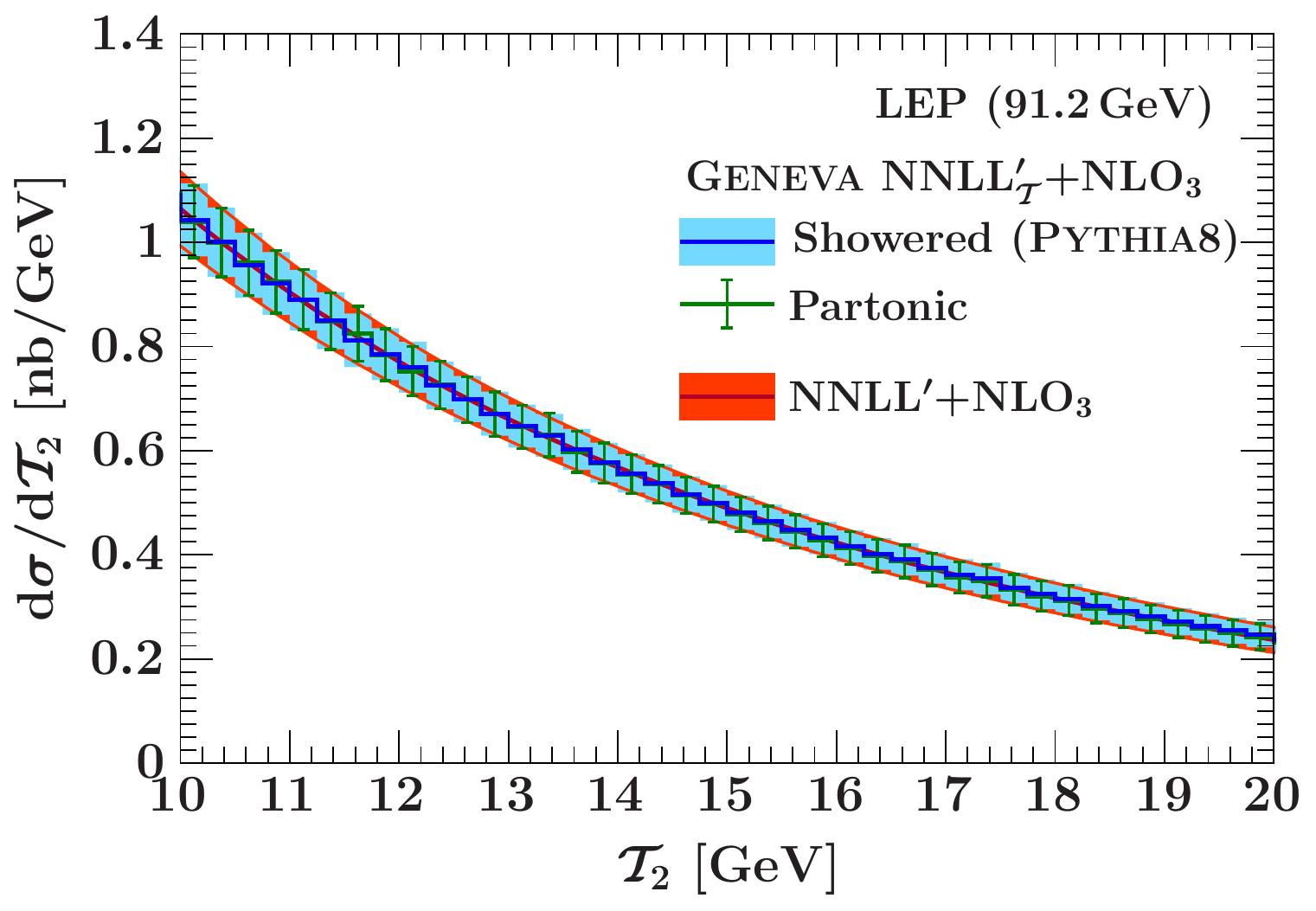}}
\end{minipage}
\begin{minipage}{0.33\linewidth}
\centerline{\includegraphics[width=\linewidth]{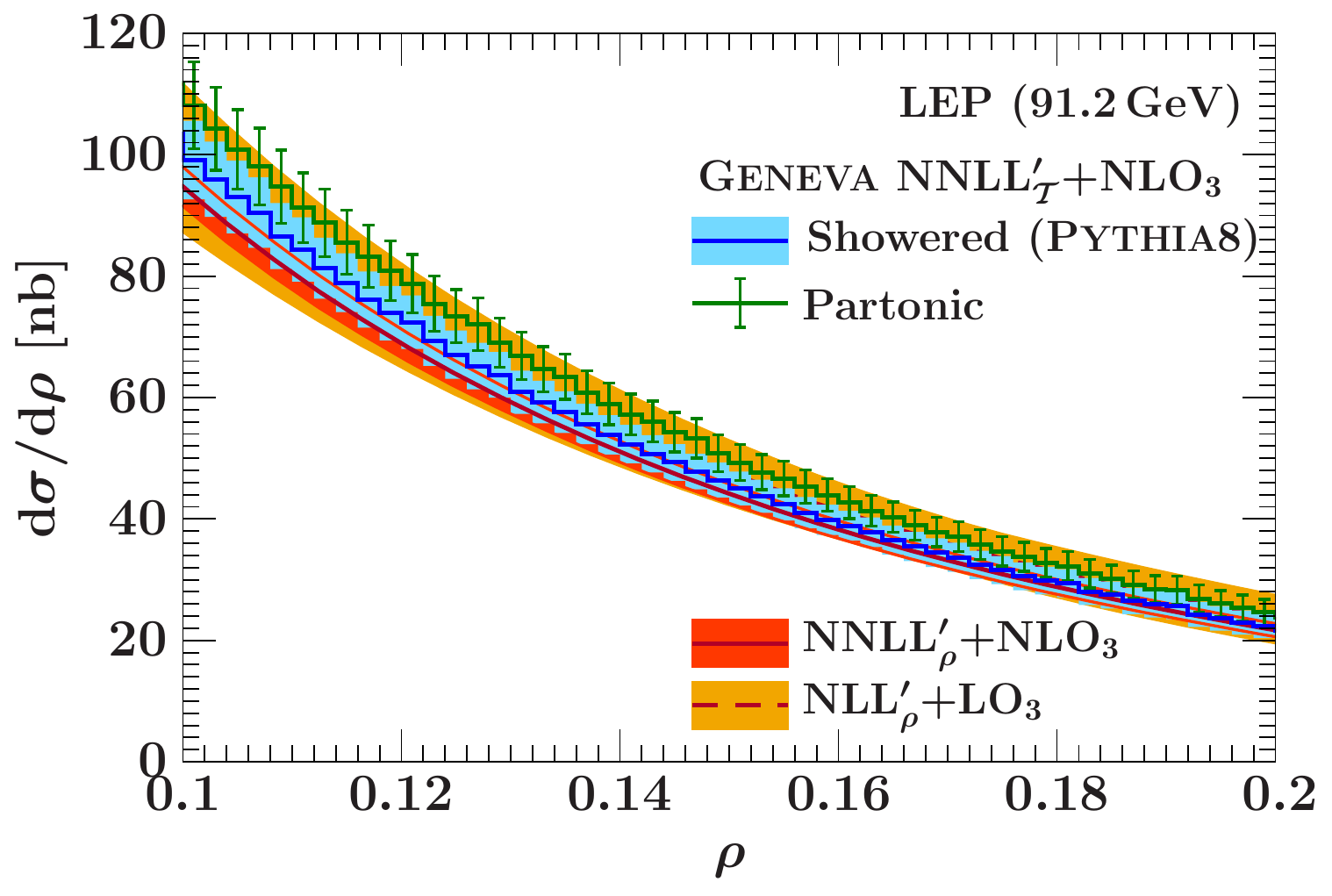}}
\end{minipage}
\begin{minipage}{0.33\linewidth}
\centerline{\includegraphics[width=\linewidth]{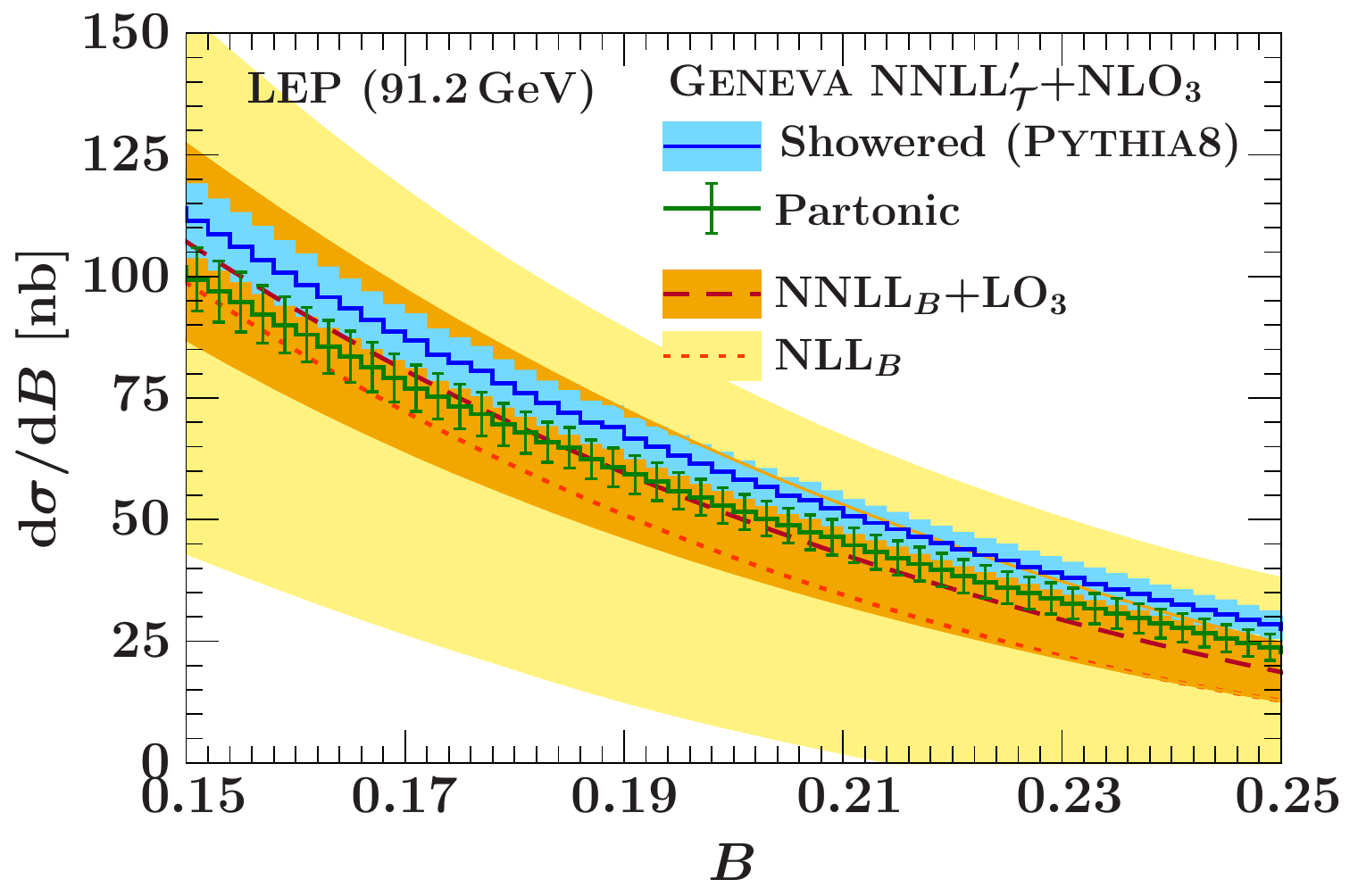}}
\end{minipage}
\caption{The $2$-jettiness (left), heavy jet mass (central), and jet broadening (right) parton-level \geneva results, compared with analytical resummation. The error bars or bands on the \geneva histograms are built from event-by-event perturbative uncertainties. Statistical uncertainties  from Monte Carlo integration are negligible. }  
\label{fig:epemDist}
\end{figure}
\begin{figure}[!ht]
\begin{minipage}{0.33\linewidth} \centerline{\includegraphics[width=\linewidth]{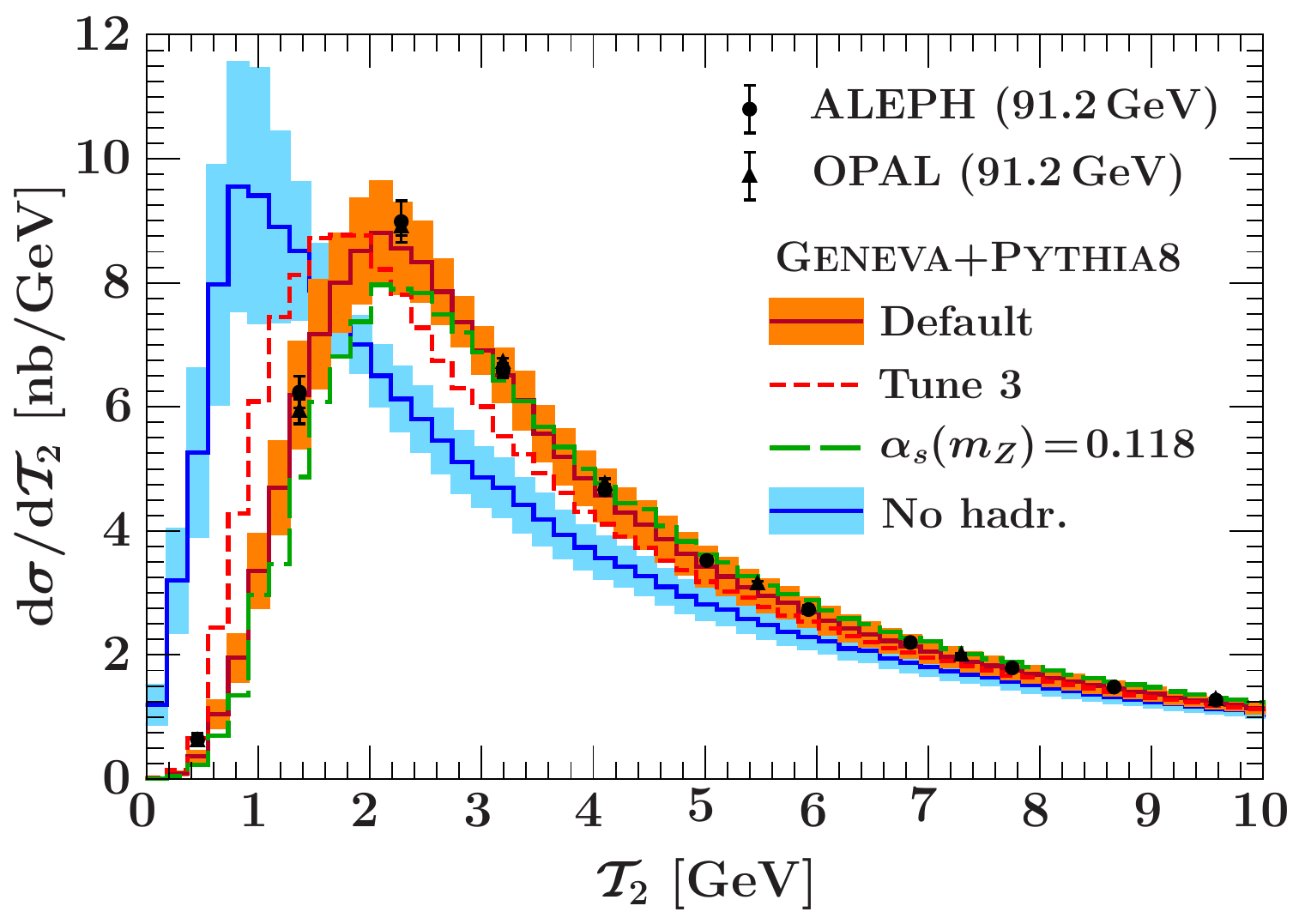}}
\centerline{\includegraphics[width=\linewidth]{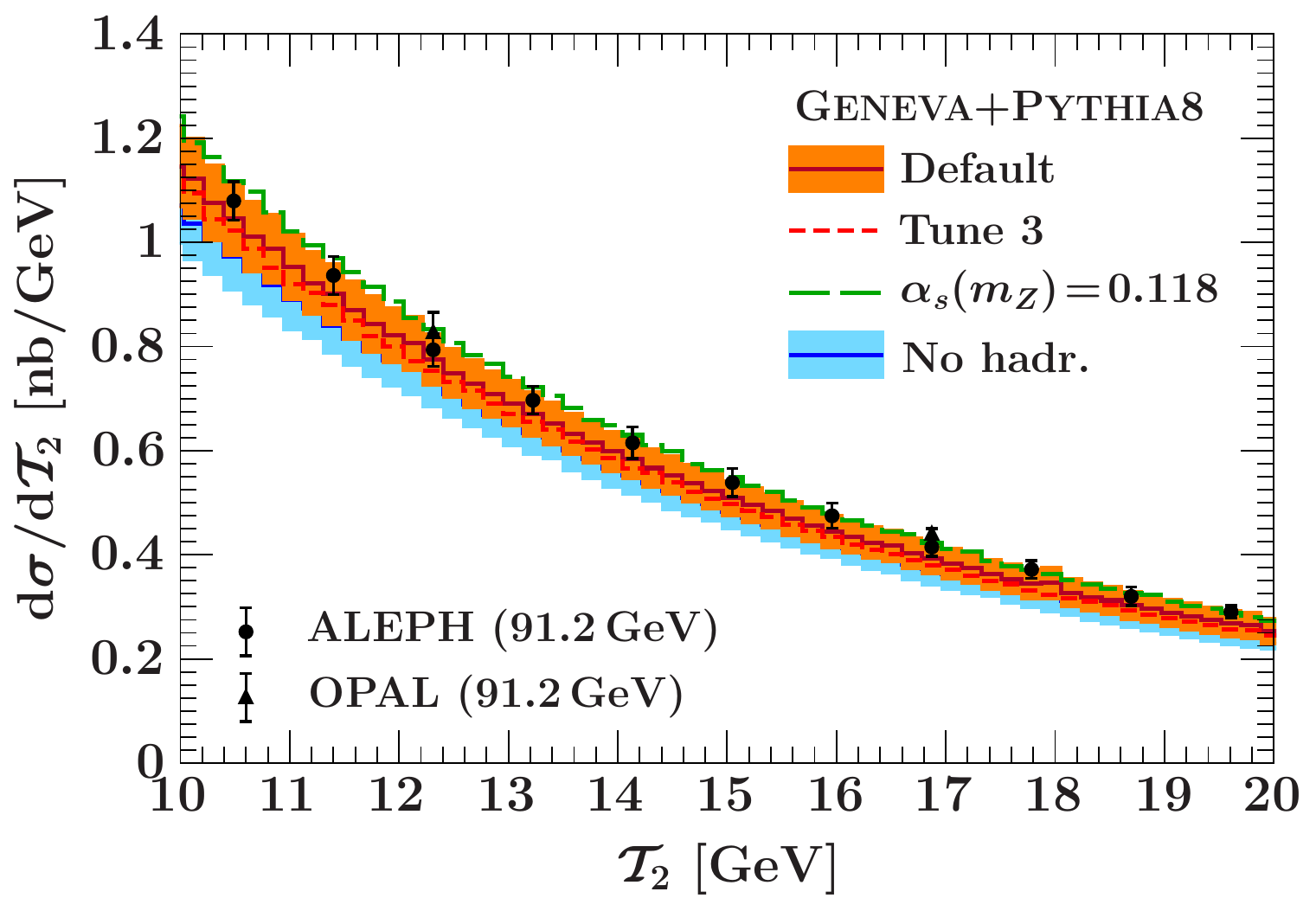}}
\end{minipage}
\begin{minipage}{0.33\linewidth}
\centerline{\includegraphics[width=\linewidth]{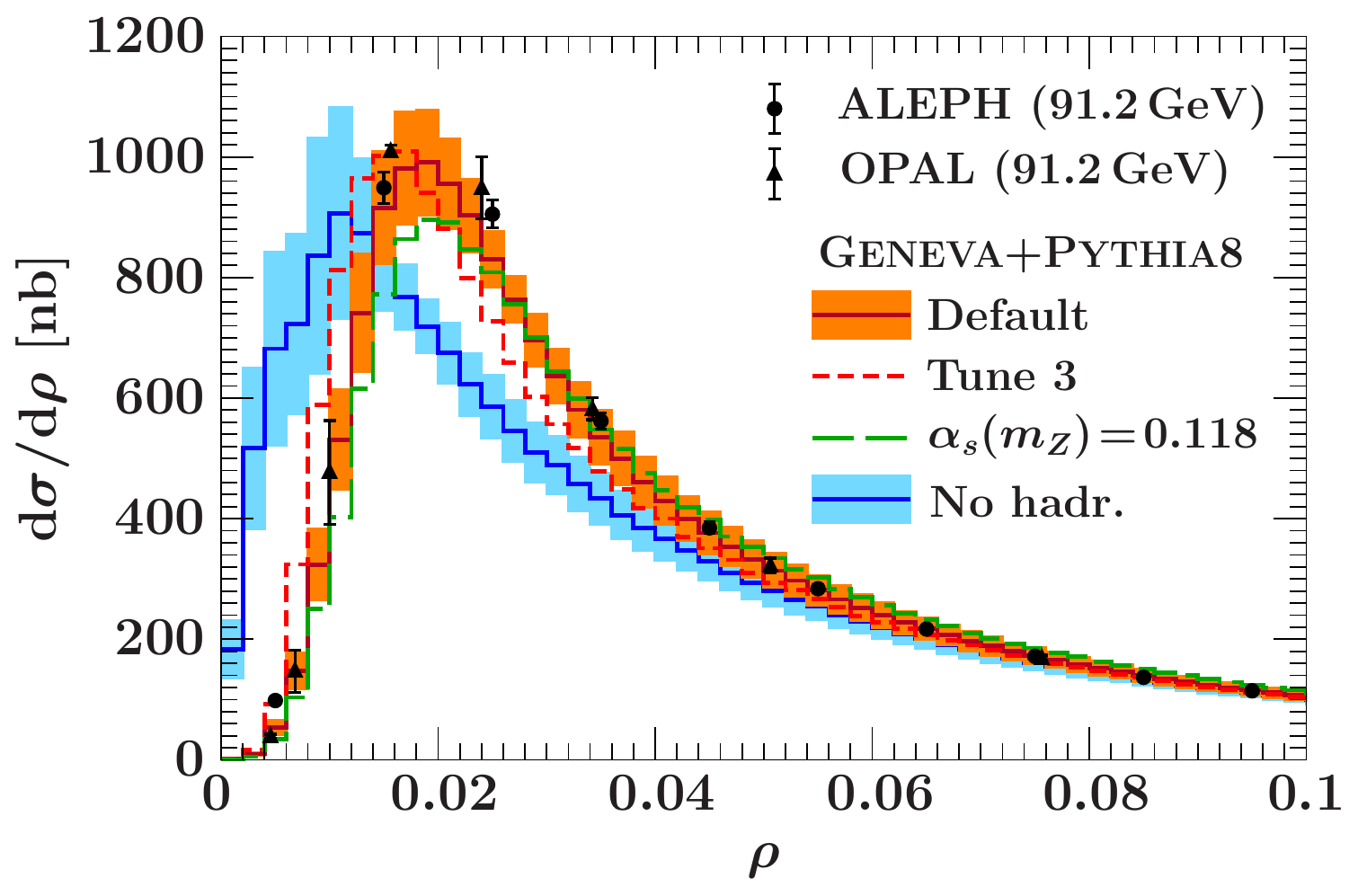}}
\centerline{\includegraphics[width=\linewidth]{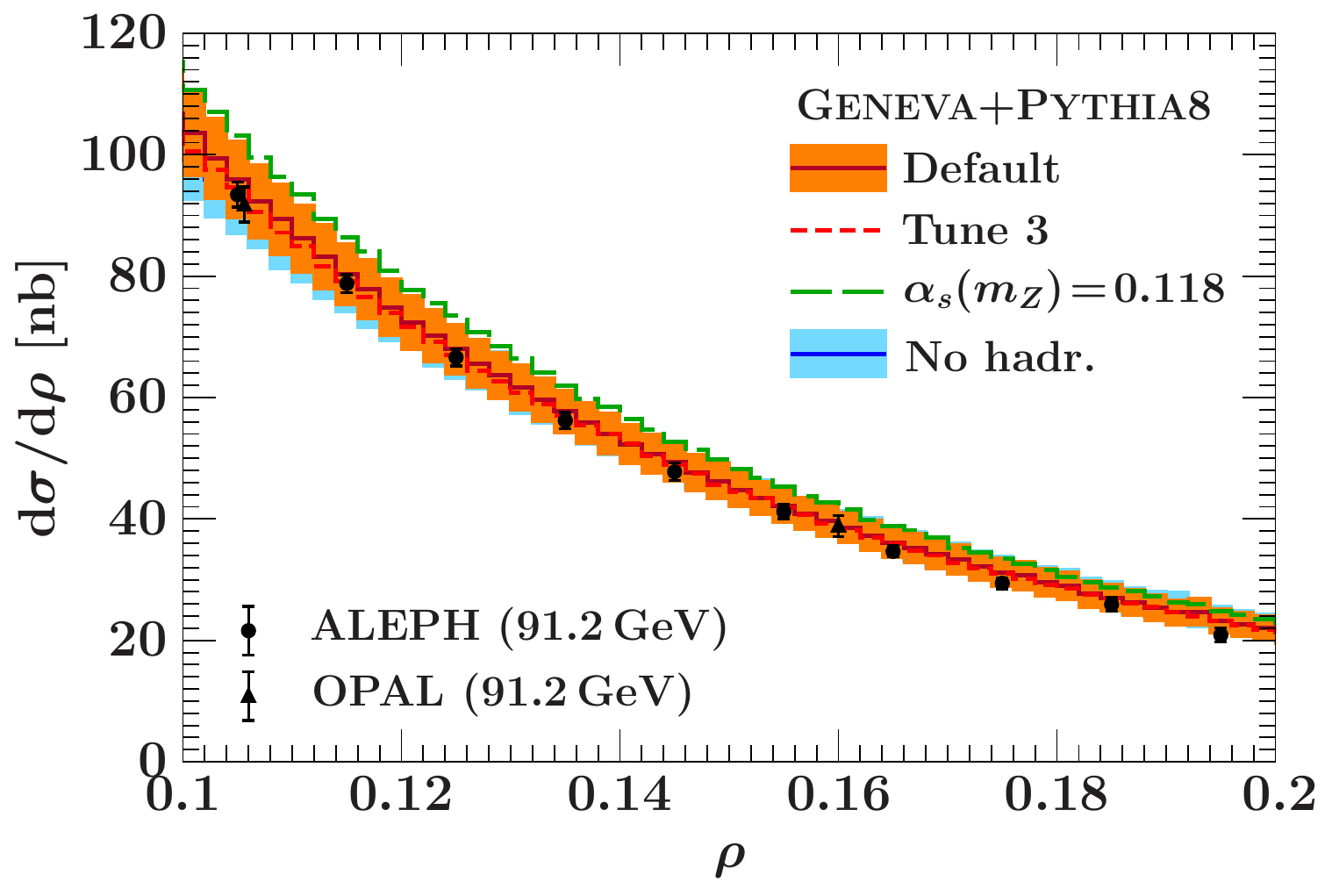}}
\end{minipage}
\begin{minipage}{0.33\linewidth}
\centerline{\includegraphics[width=\linewidth]{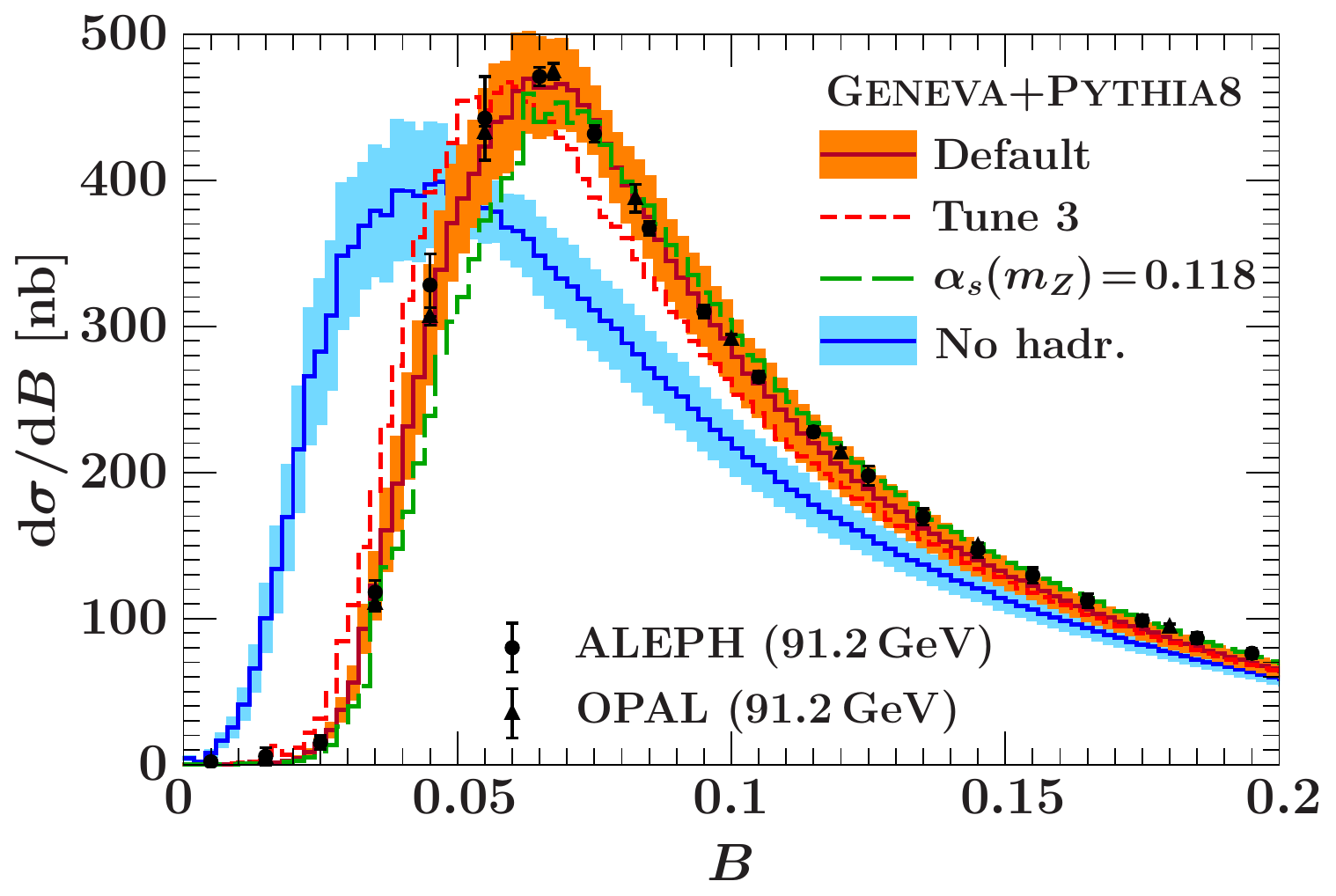}}
\centerline{\includegraphics[width=\linewidth]{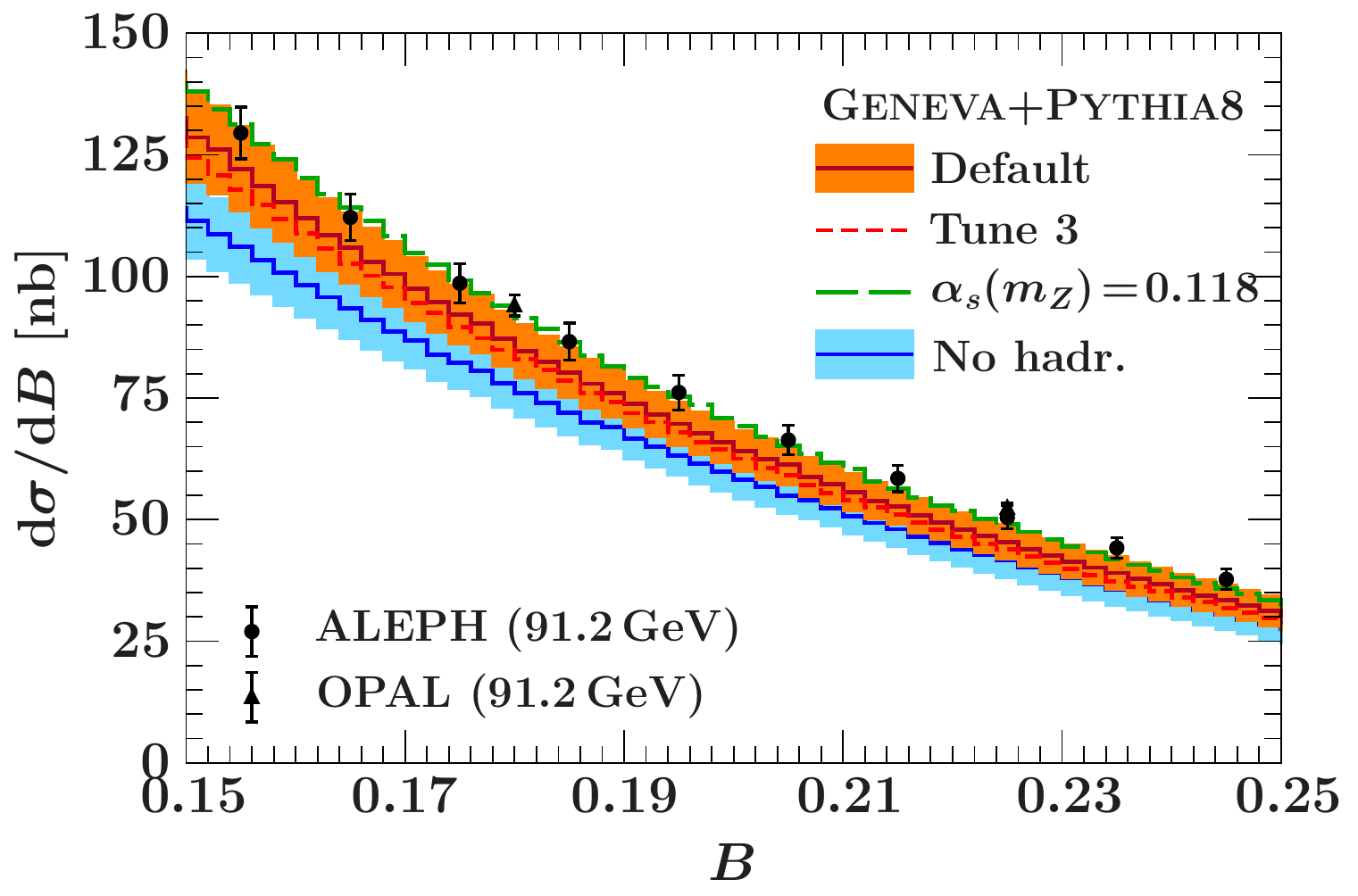}}
\end{minipage}
\caption{The $2$-jettiness (left column), heavy jet mass (central column), and  jet broadening (right column) distributions of \geneva interfaced to \pythia 8, compared to \Aleph and \Opal data in the peak (upper line) and  transition regions (lower line). Default results are obtained with \pythia 8 $e^+e^-$ tune 1 and $\alpha_s (m_Z)=0.1135$. Variations of the \pythia 8 tune, the $\alpha_s (m_Z)$ value, and results without hadronization are shown for comparison.}  
\label{fig:epemResults}
\end{figure}
We first present results for  $e^+ e^- \to 2/3$ jets, using 2-jettiness $\Tau_2$ as the $2$-jet resolution variable,
\begin{equation}\label{eq:defTau2}
\Tau_2 =\Ecm \biggl(1 - \max_{\hat n} \frac{\sum_k \abs{\hat n\cdot \vec p_k}}{\sum_k \abs{\vec p_k} }\biggr)
\,,\end{equation}
which is simply related to thrust $T$ by $\Tau_2 = \Ecm (1-T)$.  We
perform the resummation in $\Tau_2$ to NNLL$'$ and include the full
NLO$_2$, NLO$_3$, and LO$_4$ fixed-order matrix elements; i.e., we
obtain NNLL$'_\Tau$+NLO$_3$ predictions. 

 In
Fig.~\ref{fig:epemDist}, we show our results before and after \pythia8 showering, compared with analytical resummations, for $2$-jettiness, heavy jet mass, and jet broadening. We focus on the transition region, where the resummation and fixed-order calculations are
both important, and their proper combination is necessary.
\geneva results are  obtained with $\Ecm = 91.2\GeV$, $\alpha_s(m_Z) =
0.1135$ (from N$^3$LL$'$ thrust fits~\cite{Abbate:2010xh}), and \pythia 8.170 with $e^+e^-$ tune 1. The perfect agreement for the $\Tau_2$ distribution, in both the central value and in the theoretical uncertainties, is a nontrivial crosscheck
on the correctness of our implementation.
Predictions for observables other than $2$-jettiness are instead important to validate the \geneva framework, since the logarithmic structure of these observables  will in general be different from that of $2$-jettiness.
The close agreement with the analytic resummed results we find demonstrates that \geneva is able to capture a large set of higher-order logarithms for observables other than the jet resolution variable $\Tau_2$.
In Fig.~\ref{fig:epemResults}, we show our final results, including \pythia8 hadronization,
finding excellent agreement with \Aleph and \Opal data.

\begin{figure}
\begin{minipage}{0.33\linewidth}
\centerline{\includegraphics[width=\linewidth]{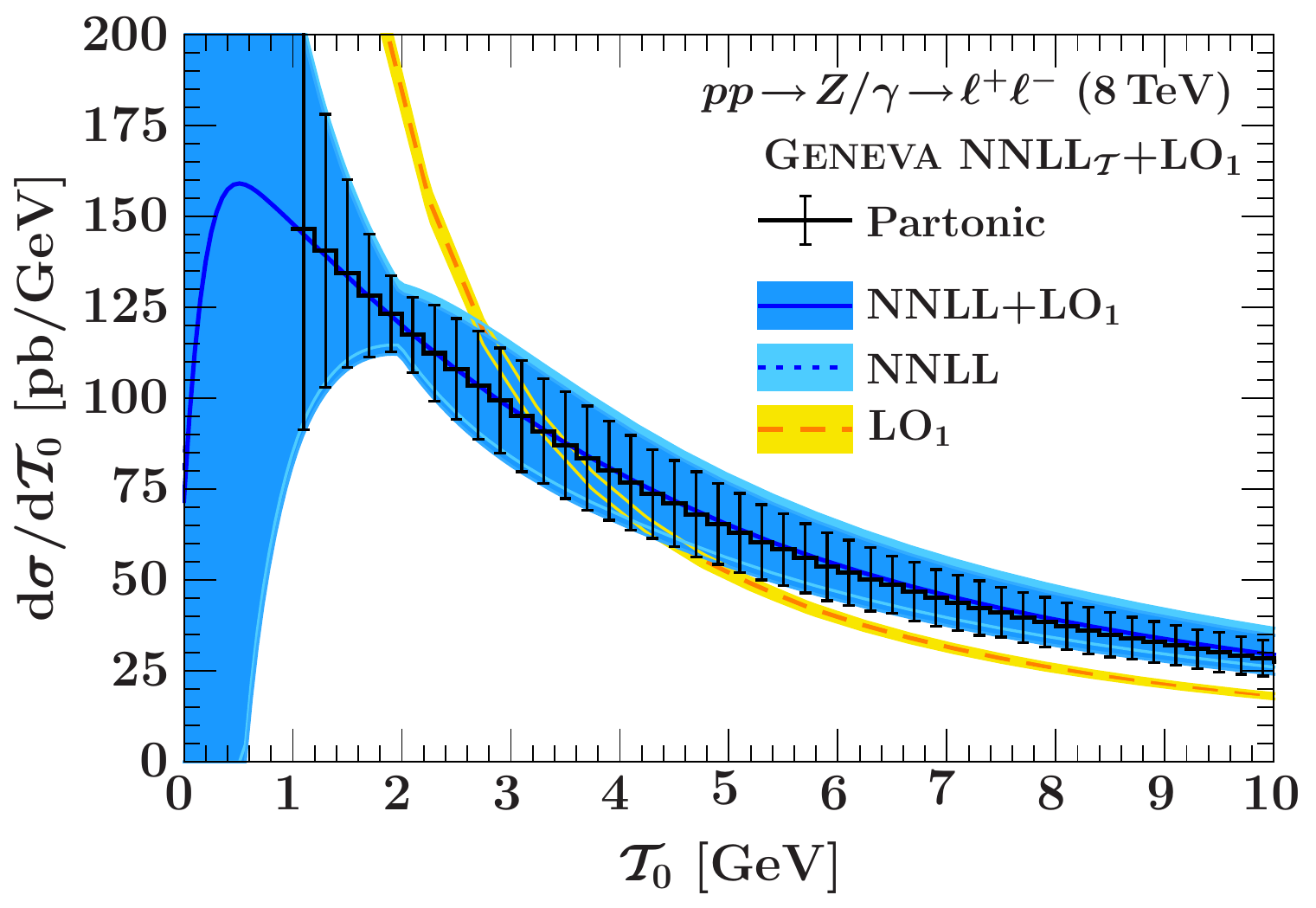}}
\end{minipage}
\begin{minipage}{0.33\linewidth}
\centerline{\includegraphics[width=\linewidth]{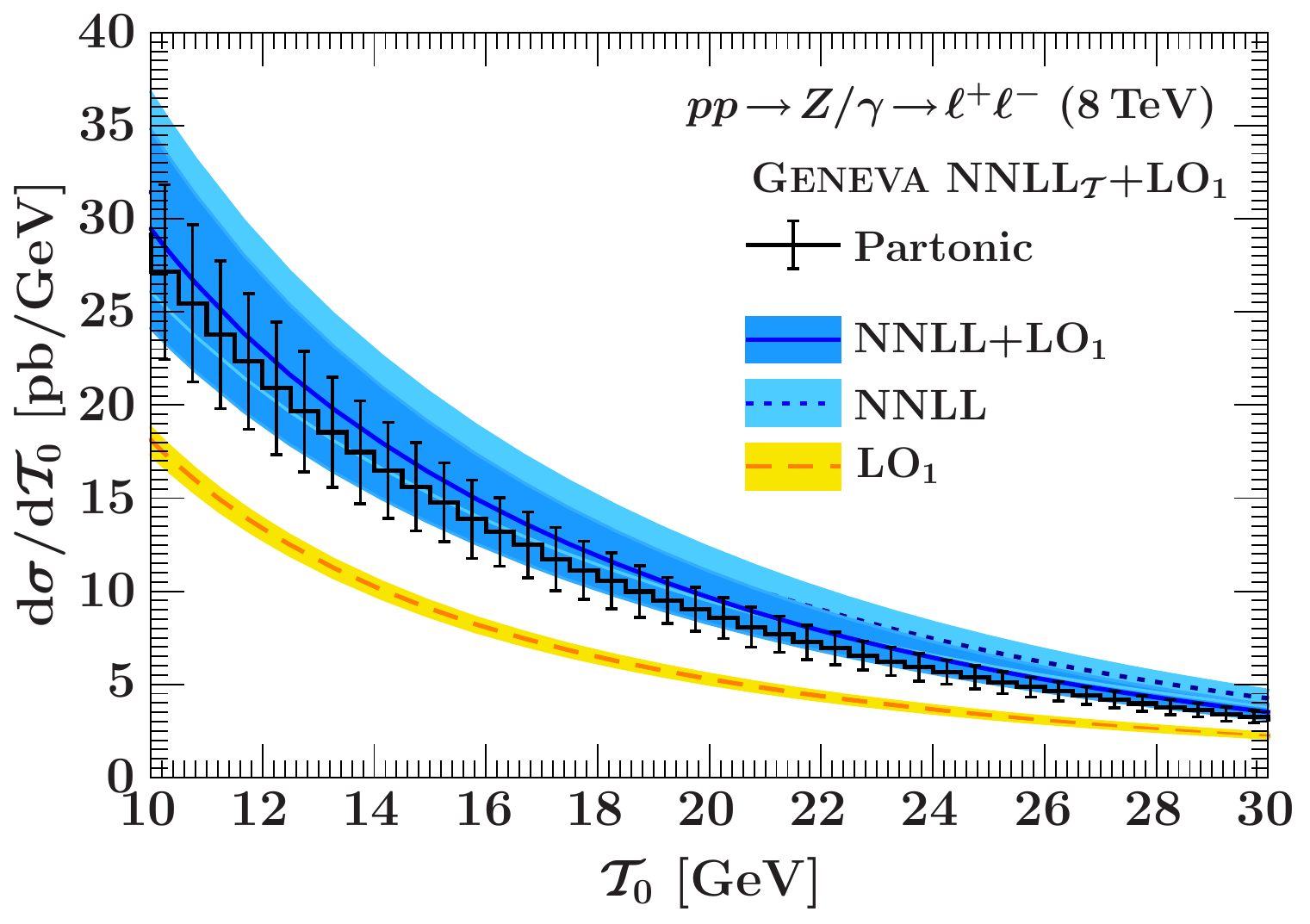}}
\end{minipage}
\begin{minipage}{0.33\linewidth}
\centerline{\includegraphics[width=\linewidth]{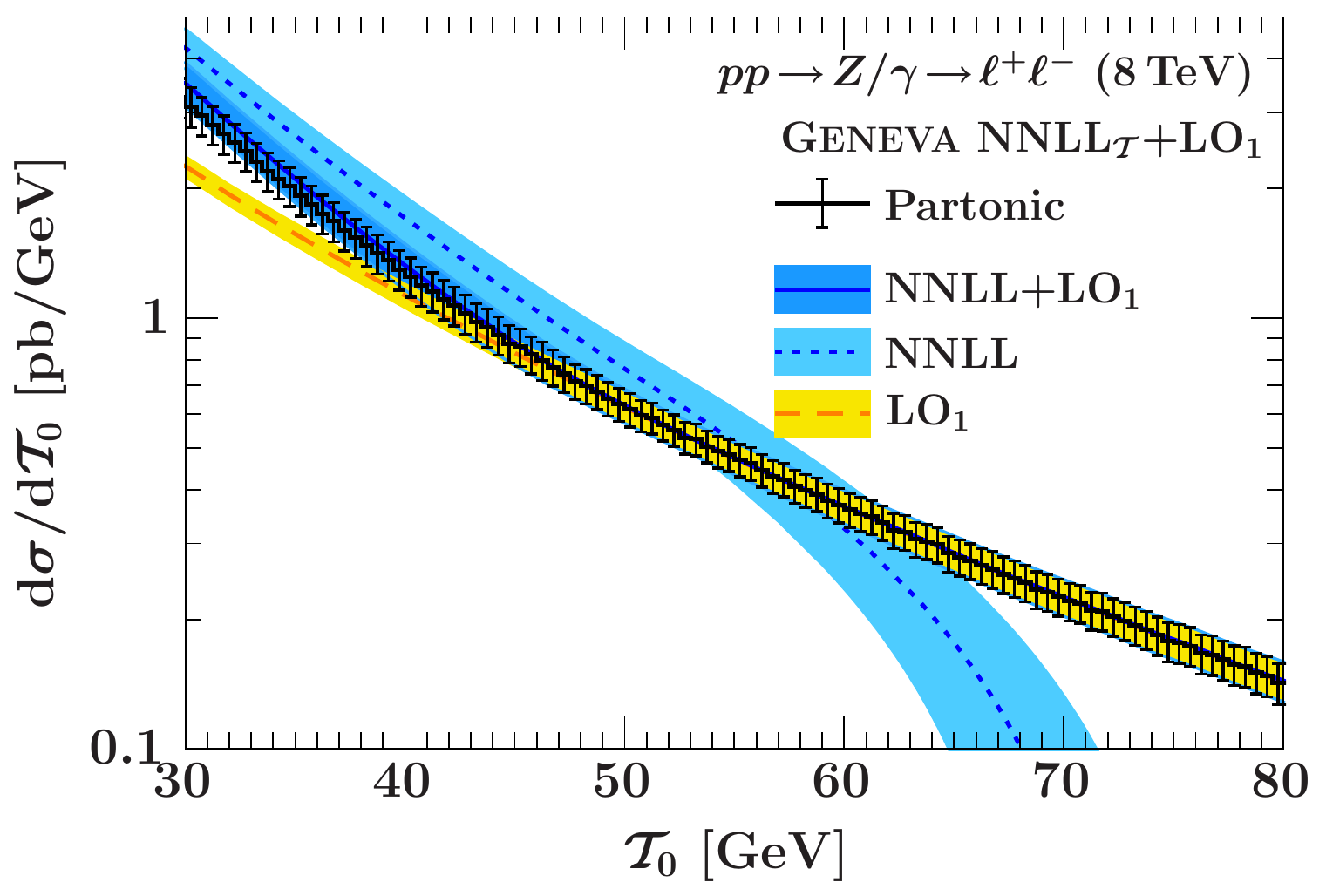}}
\end{minipage}
\caption{The \geneva partonic results for Drell-Yan production, compared to the analytic resummation of $\Tau_0$ matched to fixed order at NNLL+LO$_1$,  in the peak (left), transition (center), and tail (right) regions.}
\label{fig:Tau0Val}
\end{figure}

Next we discuss the ongoing extension to hadronic collisions. We show first results for $p p \to (Z/\gamma^* \to \ell^+ \ell^-)$ + jets, matching the 0- and 1-jet multiplicities, and using beam thrust,~\cite{Stewart:2009yx} as the resolution parameter. 
An additional complication compared to the $e^+ e^-$ case is the presence of initial-state radiation. In the resummation, the collinear radiation from the incoming partons is described by beam functions, which can be factorized into a convolution of the usual parton distribution functions and perturbatively calculable coefficients.~\cite{Stewart:2009yx}
In Fig.~\ref{fig:Tau0Val}, we show \geneva  results at NNLL+LO$_1$ for Drell-Yan production in $pp$ collisions at $\Ecm = 8$ TeV, sampling the invariant mass $Q$ of the $\ell^+ \ell^-$ pair around the $Z$ pole in the $M_Z \pm 10\ \Gamma_Z$ interval. The agreement of central values and theoretical uncertainties with the fixed-order calculation at large $\Tau_0$ and with the NNLL analytic resummation at low $\Tau_0$ serves as a useful validation.  

\section{Conclusions}
From the Monte Carlo perspective, the \geneva framework achieves the combination of higher-order resummation with multiple NLO calculations. From the resummation perspective, it allows one to obtain fully differential results that correctly include the resummation of the jet resolution variable to higher logarithmic accuracy.  

We have presented results for $e^+e^-$ collisions, employing $2$-jettiness as a resolution parameter. Using $\alpha_s(m_Z) = 0.1135$ together with tune 1 of \pythia8, we obtain an excellent description of \Aleph and \Opal data, both for thrust and for observables whose resummation structure is distinct from that of $2$-jettiness, namely $C$-parameter, heavy jet mass, and jet broadening. 

The extension to $pp$ collisions is in progress. Here, we have concentrated on the Drell-Yan process at the LHC, using beam thrust as the jet resolution variable and showing the first steps of such an implementation.

\section*{Acknowledgments}
This work was supported by  DOE grants DE-PS02-09ER09-26, DE-FGO2-96ER40956, NSF grants NSF-PHY-0705682, NSF-PHY-0969510, DFG EM grant No. TA 867/1-1 and used resources of NERSC, supported by the DOE grant DE-AC02-05CH11231.

\end{document}